\documentclass{eptcs}

\def \hvar{{\sf h}}
\def \hrvar{{\sf hr}}
\def \ttvar{{\sf tt}}
\def \outtvar{{\sf out_t}}
\def \outIvar{{\sf out_1}}
\def \outIIvar{{\sf out_2}}
\def \pctvar{{\sf pc_t}}
\def \hpctvar{{\sf \widehat{pc}_t}}
\def \hpcIvar{{\sf \widehat{pc}_1}}
\def \hpcIIvar{{\sf \widehat{pc}_2}}
\def \calltvar{{\sf call_t}}
\def \callIvar{{\sf call_1}}
\def \callIIvar{{\sf call_2}}
\def \endbox{\hfill \qed}
\providecommand{\event}{REFINE 2015} % Name of the event you are submitting to

\newcommand\Tstrut{\rule{0pt}{2.6ex}}         % = `top' strut
\newcommand\Bstrut{\rule[-0.9ex]{0pt}{0pt}}   % = `bottom' strut

\def \outt {$\mathtt{out_1}$} 
\def \outu {$\mathtt{out_2}$}
\def \follows {\Leftarrow} 

\def \nasgn {\mathbin{:\joinrel\in}}
\def \qa {\mathbin{?}}

\usepackage{graphicx}
\usepackage{rcp}
\usepackage{url}
\usepackage{amsmath,amssymb}
\usepackage{color}
\usepackage{stmaryrd}
\usepackage{oz}
\usepackage{zed-csp}

\newcommand{\rstmt}[1]{\langle #1 \rangle}

\newcommand{\beh}[1]{\llbracket #1 \rrbracket}

\newcommand{\rStut}{rS}
\newcommand{\rPriv}{rL}

\def \diverge {{\mathop \uparrow}}

\def \traces {{\sf Tr}}

\def \all  {\forall}
\def \asgn  {{\;:=\;}}

\def \refs  {{\;\sqsubseteq\;}}

\def \sref  {\sqsubseteq}

\def \lvar {{\bf var_u}\ }
\def \gvar {{\bf var_o}\ }
\def \entails {\Rrightarrow}

\newcommand{\sskip}{\mathbf{skip}}

\newcommand{\ddo}{\mathbf{do}}
\newcommand{\ood}{\mathbf{od}}

\renewcommand{\End}{{\bf end}\ }
\newcommand{\Class}{{\bf class}\ }
\newcommand{\Var}{{\bf var}\ }
\renewcommand{\Init}{{\bf initialisation}\ }
\newcommand{\Method}{{\bf method}\ }
\newcommand{\When}{{\bf when}\ }
\newcommand{\While}{{\bf while}\ }
\renewcommand{\Then}{{\bf then}\ }
\newcommand{\Do}{{\bf do}\ }
\newcommand{\Action}{{\bf action}\ }
\newcommand{\Assert}{{\bf assert}\ }
\renewcommand{\Begin}{{\bf begin}\ }
\renewcommand{\If}{{\bf if}\ }
\renewcommand{\Else}{{\bf else}\ }
\newcommand{\New}{{\bf new}\ }

\newcommand{\mcA}{\mathcal{A}}
\newcommand{\mcB}{\mathcal{B}}
\newcommand{\mcC}{\mathcal{C}}
\newcommand{\mcP}{\mathcal{P}}
\newcommand{\mcR}{\mathcal{R}} 
\newcommand{\mcT}{\mathcal{T}}

\newcommand{\mcM}{\mathcal{M}}

\newcommand{\Abort}{\mathbf{abort}}
\newcommand{\Magic}{\mathbf{magic}}

\newcommand{\refeq}[1]{(\ref{#1})}
\newcommand{\refeqn}[1]{(\ref{#1})}

\newcommand{\reffig}[1]{Figure~\ref{#1}}
\newcommand{\refthm}[1]{Theorem~\ref{#1}}
\newcommand{\reflem}[1]{Lem\-ma~\ref{#1}}

\newcommand{\refsec}[1]{Section~\ref{#1}}
\newcommand{\refex}[1]{Example~\ref{#1}}
\newcommand{\refdef}[1]{Definition~\ref{#1}}

\newcommand{\reftab}[1]{Table ~\ref{#1}}

\usepackage{amsthm}
\theoremstyle{plain}
\newcounter{thm}
\newtheorem{theorem}[thm]{Theorem}
\newtheorem{lemma}[thm]{Lemma}

\newtheorem{corollary}[thm]{Corollary}
\theoremstyle{definition}
\newtheorem{definition}[thm]{Definition}
\newtheorem{example}[thm]{Example}

\title{Towards linking correctness conditions for concurrent objects
  and contextual trace refinement}

\author{Brijesh Dongol \\
{Department of Computer Science, Brunel University London} \\
\texttt{Brijesh.Dongol@brunel.ac.uk}
\and
Lindsay Groves \\
{School of Engineering and Computer Science, Victoria University 
of Wellington} \\
\texttt{lindsay@ecs.vuw.ac.nz}\\
}

\def \authorrunning{Brijesh Dongol and Lindsay Groves} 
\def \titlerunning{Linking correctness conditions and
  contextual trace refinement}

\begin{document}

\maketitle

\begin{abstract}
  Correctness conditions for concurrent objects describe how atomicity
  of an abstract sequential object may be decomposed. Many different
  concurrent objects and proof methods for them have been developed.
  However, arguments about correctness are conducted with respect to
  an object in isolation. This is in contrast to real-world practice,
  where concurrent objects are often implemented as part of a programming
  language library (e.g., \texttt{java.util.concurrent}) and are
  instantiated within a client program. A natural question to ask, then is:
  How does a correctness condition for a concurrent object ensure
  correctness of a client 
  program that uses the concurrent object? 
% A first step towards
  % answering this question was taken by Filipovi\'{c} et al., who
  % establish a link between correctness conditions sequential
  % consistency and linearizability, and a contextual notion of
  % refinement called observational refinement. However, their results
  % are only valid for terminating client programs, i.e., cannot cope
  % with reactive clients (e.g., operating systems). 
  This paper presents the main issues that surround this question and
  provides some answers by linking different correctness conditions with a
  form of trace refinement.
  % Filipovi\'{c} et al.'s framework by linking correctness conditions
  % with contextual trace refinement, which enables reasoning about
  % non-terminating, divergent and faulty clients.
  % Within our framework
  % we focus on sequential consistency and linearizability: we show that
  % sequential consistency is inadequate for contextual trace
  % refinement, but that linearizability implies contextual trace
  % refinement for all canonical implementations of a sequential
  % specification object. The modular nature of our framework allows the
  % connections between other correctness conditions and contextual
  % trace refinement to be studied.
\end{abstract}

\section{Introduction}

Concurrent objects provide operations that may be invoked by the
parallel threads of a concurrent client, enabling more efficient
computation in, for example, modern multi-core
architectures. % \reffig{fig:client-object} depicts a multithreaded
% client program that calls operations $Op_1$ and $Op_2$ of a concurrent
% object possibly with some inputs and outputs.
A concurrent object
could be a commonly-used data structure such as a stack, queue, or
set, or could implement a novel programming paradigm such as a software
transactional memory object, which allows several reads and writes to
be treated as a single atomic transaction \cite{HeSh08}.

% \begin{figure}[ht]
%   \rule{\textwidth}{1pt}\smallskip

%   \centering 
  
%     \scalebox{0.7}{\input{client-object-2.pspdftex}}
%     \caption{Client-object relationship: The arrows represent the
%       invocations and responses between the client program and the
%       concurrent object it uses}
%     \label{fig:client-object}
%   % \begin{minipage}[b]{0.49\linewidth}
%   %   \centering
%   %   \scalebox{0.5}{\input{stack-exec-lin.pspdftex}}
%   %   \caption{Linearizable stack execution}
%   %   \label{fig:lin}
%   % \end{minipage}
%   \rule{\textwidth}{1pt}

% \end{figure}

Correctness conditions, such as linearizability \cite{Herlihy90}, for
concurrent objects cannot be defined in terms of pre/post conditions for
their operations due to the possibility of interference while the operations
are executing.  Instead, they are defined in terms of a relation between
(concurrent) histories of a concrete implementation and 
(sequential) histories of its abstract counterpart, which
records occurrence of certain events at the concrete and abstract
levels, such as invocations and responses of method calls
\cite{DDGS15-ECOOP}.

Many papers have been devoted to verifying linearizability of
concurrent object implementations \cite{DongolD14-csur}; however, these
typically only consider behaviours of the concurrent 
object at hand in isolation --- they do not provide any guarantees to the
client programs that use concurrent objects. Therefore, letting $P[O]$ denote
a client program $P$ that uses object $O$, we consider the following question:
\begin{quote}
  \emph{Provided concurrent object $OC$ is correct with respect to
    sequential object $OA$, how are the behaviours of $P[OA]$ related
    to those of $P[OC]$}?
\end{quote}
This question was been examined by Filipovi\'{c} et al.\
\cite{FORY10}, who establish a link between two correctess conditions:
\emph{sequential consistency} and \emph{linearizability} with a notion
of refinement called \emph{observational refinement}, which is
essentially \emph{data refinement} as defined by He et al.\
\cite{HeHS86}. They show that sequential consistency and observational
refinement coincide when threads are \emph{data independent} (i.e.,
there is no data shared between client threads), while linearizability
coincides with observational refinement when data dependence is allowed.

This paper provides a brief overview of our investigation, which examines
this question in a more general setting than Filipovi\'{c} et al.  A more
detailed account of this work will be published elsewhere.

% Our paper builds on Filipovi\'{c} et al.'s results and addresses some
% shortcomings of their work.  In particular, we make the following
% contributions. We develop (1) a more general theory for \emph{trace
%   refinement} of client-object systems, enabling reasoning about both
% non-terminating and terminating (e.g., reactive) contexts; (2) a more
% familiar refinement framework --- Filipovi\'{c} et al. develop and use
% a somewhat unusual framework making their results difficult to
% understand, e.g., the term \emph{simulation} is used in a non-standard
% way as a relation on histories as opposed to a method for proving
% refinement; and (3) links between different correctness conditions and
% contextual trace refinement. Our framework is modular allowing the
% link between other correctness conditions % (including those for relaxed
% % memory models
% \cite{DDGS15-ECOOP} and contextual trace refinement to be studied.

% \fbox{Add outline?}

\section{Concurrent objects and their clients}
\label{sec:concurrent-objects}

% In this section, for tangibility, we present an example concurrent
% object: the non-blocking stack by Treiber in
% \refsec{sec:exampl-treib-stack}, which we use informally to discuss
% different correctness conditions. % Example stack clients are given in
% % \refsec{sec:concurrent-clients}, which we use to informally motivate
% % contextual trace refinement. % Correctness conditions and trace
% % % refinement are formalised in Sections \ref{sec:refin-client-object}
% % % and \ref{sec:trace-refin-client}, respectively.

% \subsection{Example concurrent object: The Treiber stack}
% \label{sec:exampl-treib-stack}
% Without garbage collection, solving this requires
% additional complexities such as version numbers for pointers to be
% introduced; such details are elided in this paper.
\begin{figure}[t]

  \begin{minipage}[b]{0.58\columnwidth}
    \smallskip
    
    {\tt \small Init: Head = null}\smallskip

    \begin{minipage}[t]{0.45\columnwidth}
      \tt \small push(v) ==
      
       H1:\      n := \textbf{new}(Node);
      
       H2:\ 
      n.val := v;
      
      \ \ \ \ \textbf{repeat}
      
       H3:\ 
      \ \ ss := Head;
      
       H4:\ 
      \ \ n.next := ss;
      
       H5:\ 
      \textbf{until} 

      \ \ \ \ \ \ CAS(Head,ss,n)
      
       H6:\ 
      \textbf{return}
    \end{minipage}
    \begin{minipage}[t]{0.51\columnwidth}
      \tt \small pop ==
      
       \ \ \ \
      \textbf{repeat }
      
       P1: 
      \ ss := Head;
      
       P2: 
      \ \textbf{if} ss = null 
      
      P3: \ \ \ \textbf{then return} empty
      
      % \ \ \ \ \ \ \ \textbf{endif};
      
       P4: 
      \ ssn := ss.next;
      
       P5: 
      \ lv := ss.val
      
       P6: 
      \textbf{until} 
      
      \ \ \ \ \ \ \ CAS(Head,ss,ssn);
      
       P7: 
      \textbf{return} lv
    \end{minipage}
    \caption{The Treiber stack}
    \label{fig:TS}
  \end{minipage}
  \hfill 
  \begin{minipage}[b]{0.4\columnwidth}
    {\tt \small Init: S = $\emptyseq$}\medskip

    {\tt \small push(v) == \textbf{atomic} \{ S := $\langle$v$\rangle \cat$S\ \}}\medskip

    {\tt \small pop ==
      
      \ \textbf{atomic} \{ 
      
      \ \ \ \textbf{if} S = $\emptyseq$ \textbf{then}

      \ \ \ \ \ \ \textbf{return} empty
      
      \ \ \ \textbf{else} lv := head(S) ; 

      \ \ \ \ \ \ S := tail(S) ;

      \ \ \ \ \ \ \textbf{return} lv 
      
      \ \}
    }
    \caption{Abstract stack specification}
    \label{fig:Abstract-TS}
  \end{minipage}
\end{figure}

\reffig{fig:TS} presents a simplified version of a non-blocking stack
example due to Treiber \cite{Tre86}, which has become a standard case
study from the literature.\footnote{We assume that garbage
  collection is used --- this avoids the so-called ABA problem,
  where modifications to a shared pointer may go undetected when to the
  value changes from some value $A$ to another value $B$ then back to $A$.}
The implementation has fine-grained atomicity, and
each line of the {\tt push} and {\tt pop} operations corresponds to a
single atomic step. Synchronisation is achieved using an atomic
compare-and-swap (\texttt{CAS}) instruction, which takes as input a
\emph{(shared) variable} {\tt gv}, an \emph{expected value} {\tt lv}
and a \emph{new value} {\tt nv}:\medskip
%% LG: Note that new is not really atomic, but is considered atomic for our
%% purposes here.

\begin{minipage}[t]{0.9\columnwidth}
  \tt CAS(gv, lv, nv) $\sdef$ atomic \{
  \begin{tabular}[t]{@{}l@{}}
    \tt if gv = lv then gv := nv ; return true \\
    \tt else return
    false \}
  \end{tabular}
\end{minipage}\medskip

% \noindent In a single atomic step, the \texttt{CAS} operation compares
% value of {\tt gv} to {\tt lv}, potentially updates {\tt gv} to {\tt
%   nv} and returns a boolean. Specifically, if {\tt gv = lv}, it
% updates {\tt gv} to {\tt nv} and returns $true$ (to indicate that the
% update was successful), otherwise it leaves everything unchanged and
% returns $false$. The \texttt{CAS} instruction is natively supported by
% most mainstream architectures (e.g., x86).

% % \fbox{Make more general?}
% Operations that use \texttt{CAS} typically
% have a try-retry structure with a loop that stores (shared variable)
% {\tt gv} locally in {\tt lv}, performs some calculations on {\tt lv}
% to obtain {\tt nv} (a new value for {\tt gv}), then uses a
% \texttt{CAS} to attempt an update to {\tt gv}. If the \textrm{CAS}
% fails, there must have been some interference on {\tt gv} since it was
% stored locally at the start of the loop, and in this case the
% operation retries by re-reading {\tt gv}. Both the {\tt push} and {\tt
%   pop} operations of Treiber's stack have a try-retry structure, e.g.,
% in the {\tt push} the global {\tt Head} is stored locally at {\tt H3},
% and an attempt is made to update {\tt Head} at {\tt H5}.

The Treiber stack implements the abstract stack specification in
\reffig{fig:Abstract-TS}, where `$\langle$' and `$\rangle$' delimit
sequences, `$\emptyseq$' denotes the empty sequence, and `$\cat$'
denotes sequence concatenation. 
% We assume sequences are indexed from $0$ onwards. 
The abstract stack consists of a sequence of elements $S$
together with two operations $push$ and $pop$. Note that when the
stack is empty, $pop$ returns a special value {\tt empty} that cannot
be pushed onto the stack.

A correctness condition is a relationship between the \emph{histories}
of the concrete and abstract systems. Each history records the
interactions between a client and its objects. Typically, these are
invocation and return events of operation calls, which form the
object's external interface. Concurrent histories may consist of both
overlapping and non-overlapping operation calls, inducing a partial
order on events. Correctness conditions define how, if at all, this
order is maintained in the corresponding abstract history. There are
several well-known existing correctness conditions \cite{HeSh08}. In
this paper, we study two of these in detail: sequential consistency
and linearizability.
\begin{itemize}
\item \emph{Sequential consistency} \cite{Lam79} is a simple condition
  requiring the order of operation calls in a concrete history for a
  single process to be preserved. Operation calls performed by
  different processes may be reordered in the abstract history even if
  the operation calls do not overlap in the concrete history.
\item \emph{Linearizability} \cite{Herlihy90} strengthens sequential
  consistency by requiring the order of non-overlapping operations to
  be preserved. Operation calls that overlap in the concrete history
  may be reordered when mapping to an abstract history.
\end{itemize}

\paragraph{Concurrent clients.}
\label{sec:concurrent-clients}
% instead they are instantiated within a \emph{client} program which
% executes by calling the object's operations.

Correctness conditions are usually defined in terms of a \textit{most
general client} which characterises the allowable behaviours of a
concurrent object; however, they do not allow us to reason about specific
clients that use these concurrent objects.
\begin{example}
  \label{ex:1}
  The program below consists of threads {\tt 1} and {\tt
    2}, a shared stack {\tt s}, and shared variables {\tt x}, {\tt y}
  and {\tt z}.  Thread {\tt 1} pushes {\tt 1} then {\tt 2} onto 
  {\tt s}, then pops the top element of {\tt s}, and stores it
  in {\tt x}. Concurrently, thread {\tt 2} pops the top element of
  {\tt s} and stores it in {\tt y}, then reads the value of
  {\tt x} and stores it in {\tt z}.
  \begin{center}
    \begin{minipage}[t]{0.9\linewidth}
      \tt Init x, y, z = 0, 0, 0;
      \\
      \begin{minipage}[t]{0.4\columnwidth}
        \tt Thread 1:
      
        \ \ T1: s.push(1);
      
        \ \ T2: s.push(2);
      
        \ \ T3: x := s.pop();
      
      \end{minipage}
      \hfill
      \begin{minipage}[t]{0.4\columnwidth}
        \tt Thread 2:
      
        \ \ U1: y := s.pop();
      
        \ \ U2: z := x;
      \end{minipage}
    \end{minipage}
  \end{center}
  \noindent The program executes by \emph{interleaving} the atomic
  statements of the two threads.
%\footnote{For simplicity, we
%    implicitly assume an absence of race conditions.} 
  In addition,
  depending on the implementation of {\tt s}, we will get different
  behaviours of the client program because the effects of the
  concurrent operations on {\tt s} may appear to occur in different
  orders. For example, {\tt s} could be an instance of the Treiber
  Stack (\reffig{fig:TS}) (which is linearizable with respect to
  \reffig{fig:Abstract-TS}), or some other stack that satisfies a
  different correctness condition (e.g., quiescent consistency
  \cite{HeSh08}) with 
  respect to the abstract stack in \reffig{fig:Abstract-TS}.

  % Unlike the statements in \reffig{fig:TS}, the statements of client
  % threads in \refex{ex:1} are not necessarily atomic because they may
  % invoke non-atomic operations. For example, if {\tt s} is an instance
  % of the Treiber stack then operation calls at, say {\tt T1} and {\tt
  %   U1}, may overlap. It also appears (in the program above) as if the
  % return of {\tt s.pop()} and the writes to {\tt x} (at {\tt T3}) and
  % {\tt y} (at {\tt U1}) are a single atomic step. However, the reality
  % at a lower-level of abstraction is different --- the return and
  % update events are different atomic steps. To make this atomicity
  % distinction explicit, we often write programs as follows, where
  % ${\tt out_t}$ is a special variable local to thread {\tt t} that
  % stores the value returned by an operation call. Variable ${\tt
  %   out_t}$ is not a real program variable; it is used for reasoning
  % purposes only.

  % \begin{center}
  %   \begin{minipage}[t]{0.9\linewidth}
  %     \tt Init x, y, z = 0, 0, 0
  %     \\
  %     \begin{minipage}[t]{0.4\columnwidth}
  %       \tt Thread 1:
        
  %       \ \ \ T1:\ s.push(1);
        
  %       \ \ \ T2:\ s.push(2);
        
  %       \ \ \ T3:\ \outt := s.pop();
        
  %       \ \ T4:\ x := \outt ;
        
  %     \end{minipage}
  %     \hfill
  %     \begin{minipage}[t]{0.4\columnwidth}
  %       \tt Thread 2:
        
  %       \ \ \ U1:\ \outu := s.pop();
        
  %       \ \ U1':\ y := \outu ;
        
  %       \ \ \ U2:\ z := x;
  %     \end{minipage}
  %   \end{minipage}
  % \end{center}  
\end{example}

With an example client program in place, we now return to the main
question for this paper: How does one judge correctness of a system
consisting of both a client and the objects it uses? More specifically,
how does a correctness condition guaranteed by a concurrent object
that a client uses affect the behaviour of the client itself?
We will also consider what sorts of client behaviours should be examined.
%; are divergence, termination and reactivity important properties? 
We address these issues as follows:

\begin{itemize}
\item First, we pin down the aspects of the system that are visible to
  an external observer. Following Filipovi\'{c} et al.\ \cite{FORY10},
  we take the observable state to be the state of the client
  variables, and the unobservable state to be the state of the objects
  they use. Therefore for the program in \reffig{ex:1}, variables {\tt
    x}, {\tt y} and {\tt z} are observable, but none of the variables
  of the stack implementation {\tt s} are observable. This allows us
  to reason about a client with respect to different implementions of
  {\tt s}.
\item Second, we determine \emph{when} a system may be observed. Unlike
  Filipovi\'{c} et al.\ \cite{FORY10}, who only observe the state at
  the beginning and end of a client's execution, we take the states
  \emph{throughout} a client's execution to be visible. This 
  allows us to accommodate, for example, reactive or interactive
  clients, which may not terminate.
\end{itemize}

Taking both issues into account, our notion of correctness for the
combined system will be a form of \emph{contextual refinement}, which
holds iff every (observable) trace of a client that uses a concurrent
object is equivalent to some (observable) trace of the client using the
abstract specification. 
We say that $TS$ \emph{contextually trace refines} $AS$ \emph{with
  respect to the client program $P$} (denoted $AS \sref_{P} TS$) iff
every trace of $P[TS]$ is a possible trace of $P[AS]$. In this paper,
we wish to know not only whether there is an abstract client trace
equivalent to every concrete client trace, but also whether contextual
refinement holds for every client program. To this end, we say $TS$
\emph{contextually trace refines} $AS$ (denoted $AS \sref TS$) iff
$TS$ contextually trace refines $AS$ with respect to every client
program $P$.% $AS \sref_M TS$ holds for \emph{every} client program
% $M$.

% Both $AS \sref_{L} TS$ and $AS \sref TS$ are defined formally in the
% next section.

\section{Linking correctness and contextual trace refinement}
\label{sec:link-corr-cont}

We now use the framework from the previous sections to explore the
links between some well-known correctness conditions and contextual trace
refinement. % We consider sequential consistency in
% \refsec{sec:sequ-cons} and linearizability in
% \refsec{sec:linearizability}.

% Suppose $N$ is the client object above, $OA$ is the abstract stack
% specification (given in \reffig{fig:Abstract-TS}) and $OC$ is a
% stack implementation (code not shown), where $OC \models_{OA}
% SC$. Then $\mcC_{N, OA} \not\sref \mcC_{N,OC}$. 

\paragraph{Sequential consistency.}
\label{sec:sequ-cons} 
Our main result for sequential consistency and contextual trace
refinement is negative --- sequential consistency does not guarantee
contextual trace refinement of the underlying clients, regardless of
whether the client program in question is data independent.

\begin{lemma}
  \label{lem:sequ-cons-2}
  Suppose $N$ is a client object and $OA$, $OC$ are concurrent objects
  such that $OC$ is sequentially consistent with respect to $OA$. Then
  it is not necessarily the case that $N[OA] \sref N[OC]$ holds.
\end{lemma}
% The proof is via Examples \ref{ex:sc1} and \ref{ex:sc2}, which provide
% counter-example instantiations of $N$, $OA$ and $OC$, covering clients
% with and without data sharing.

% [Sequential consistency does not guarantee trace refinement for data
% dependent clients]
% \begin{example}
%   \label{ex:sc1}
%   Suppose {\tt s} is a sequentially consistent stack operated on by a
%   client with two threads ${\tt 1}$ and ${\tt 2}$, which also share
%   variables ${\tt x}$ and ${\tt y}$. 
%   \begin{center}
%     \begin{minipage}[t]{0.8\linewidth}
%       \tt Init x, y = 0
      
%       \begin{minipage}[t]{0.4\columnwidth}
%         \tt Thread 1 ==
        
%         \ \ \ T1:\ s.push(1);
        
%         \ \ \ T2:\ s.push(2);
        
%         \ \ \ T3:\ \outt\ \asgn s.pop();

%         \ \ T4:\ x \asgn \outt;
        
%       \end{minipage}
%       \hfill
%       \begin{minipage}[t]{0.4\columnwidth}
%         \tt Thread 2 ==
        
%         \ \ \ U1:\ await(x != 0);
        
%         \ \ \ U2:\ \outu\ \asgn s.pop();

%         \ \ U3:\ y \asgn \outu;
%       \end{minipage}
%     \end{minipage}
%   \end{center}
%   At the concrete level, sequential consistency would hold even if the
%   \texttt{pop} at line {\tt T3} returns $1$, and the \texttt{pop} at
%   line {\tt U1} returns $2$, resulting in the (observable) trace:
%   $\langle (x, y) \mapsto (0, 0), \ \ (x, y) \mapsto (1,0), \ \ (x, y)
%   \mapsto (1,2)\rangle$, where $(x, y) \mapsto (0, 0)$ is shorthand
%   for $\{x \mapsto 0, y \mapsto 0\}$. This trace is not possible if
%   {\tt s} is the abstract stack $AS$ from
%   \reffig{fig:Abstract-TS}.\hfill \qed
% \end{example}

\begin{example} 
  \label{ex:sc2}

  Consider the program below, where the client threads are data
  independent --- {\tt x} is local to thread {\tt 1}, while {\tt y} and
  {\tt z} are local to thread {\tt 2} --- and $s$ is assumed to be
  sequentially consistent.
  \begin{center}
    \begin{minipage}[t]{0.8\linewidth}
      \tt Init x, y, z = 0;
      
      \begin{minipage}[t]{0.4\columnwidth}
        \tt Thread 1 ==
        
        \ \ \ T1:\ s.push(1);
        
        \ \ \ T2:\ s.push(2);
        
        \ \ \ T3:\ \outt\ \asgn s.pop();
        
        \ \ \ T4:\ x \asgn \outt;
        
      \end{minipage}
      \hfill
      \begin{minipage}[t]{0.4\columnwidth}
        \tt Thread 2 ==
        
        \ \ \ U1:\ z \asgn 1;
        
        \ \ \ U2:\ \outu\ \asgn s.pop();
        
        \ \ \ U3:\ y \asgn \outu;
      \end{minipage}
    \end{minipage}
  \end{center}\smallskip
  Suppose thread {\tt 1} is executed to completion, and then thread
  {\tt 2} is executed to completion. Because {\tt s} is sequentially
  consistent, the first {\tt pop} (at {\tt T3}) may set $\outIvar$ to
  {\tt 1}, the second (at {\tt U2}) may set $\outIIvar$ to {\tt
    2}. This gives the execution $ \langle (x, y, z) \mapsto (0, 0,
  0), \ \ (x, y, z) \mapsto (1,0,0), \ \ (x, y, z) \mapsto (1,0,1), \
  \ (x, y, z) \mapsto (1,2,1)\rangle $, which cannot be generated when
  using the abstract stack $AS$ from \reffig{fig:Abstract-TS} for {\tt
    s}. \hfill \qed
\end{example}

  \reflem{lem:sequ-cons-2} differs from the results of Filipovi\'{c}
  et al.\ \cite{FORY10}, who show that for data independent
  clients, sequential consistency implies observational refinement. In
  essence, their result holds because observational refinement only
  considers the initial and final states of a client program --- the
  intermediate states of a client's execution are ignored. Thus,
  internal reorderings due to sequentially consistent objects have no
  effect when only observing pre/post states. One can develop hiding
  conditions so that observational refinement is treated as a special
  case of contextual trace refinement, allowing one to obtain a
  positive result for sequential consistency equivalent to the result
  by Filipovi\'{c} et al. Full development of this theory is left for
  future work.

\paragraph{Linearizability.}
\label{sec:linearizability}

We now consider the link between linearizability and contextual trace
refinement.

% \begin{example}
%   It is straightforward to see that trace refinement holds for the
%   program in \refex{ex:sc1}. Because {\tt s} is linearizable the two
%   pops cannot occur out-of-order in any concrete trace as they cannot
%   overlap. Therefore, one will always be able to find a corresponding
%   abstract trace.  \hfill \qed
% \end{example}

\begin{example}
  Consider the program in \refex{ex:sc2}, but now assume that the
  stack {\tt s} is linearizable, e.g., is the Treiber stack.
  Reasoning that the traces generated by this program are valid (i.e.,
  have a corresponding abstract trace) requires case analysis.
%: enumerating these is straightforward. 
  In the final state
  we have either {\tt x = 1} or {\tt x = 2}. For traces that end with
  {\tt x = 1}, {\tt U2} must linearize before {\tt T3}, but after {\tt
    T2}. Therefore, {\tt U1} also occurs before {\tt T3}. For traces
  that end with {\tt x = 1}, either {\tt U2} linearizes before {\tt
    T1} or {\tt T3} linearizes before {\tt U2}. The graph below
  depicts all possible traces of the client using the concrete
  implementation.\medskip
  
  % \begin{center}
  \hfill \scalebox{0.8}{\begin{picture}(0,0)%
\includegraphics{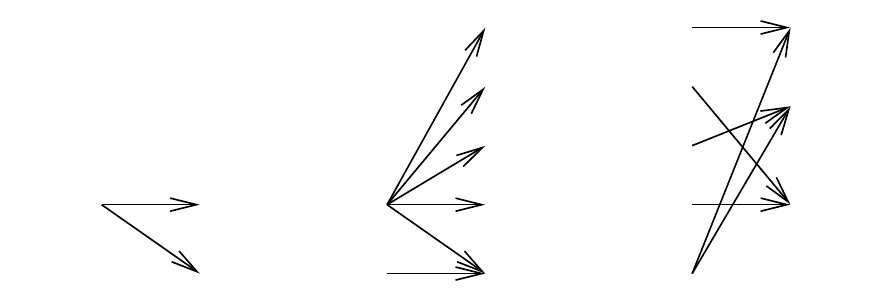}%
\end{picture}%
\setlength{\unitlength}{4144sp}%
\begingroup\makeatletter\ifx\SetFigFont\undefined%
\gdef\SetFigFont#1#2#3#4#5{%
  \reset@font\fontsize{#1}{#2pt}%
  \fontfamily{#3}\fontseries{#4}\fontshape{#5}%
  \selectfont}%
\fi\endgroup%
\begin{picture}(4080,1353)(2236,-1345)
\put(2251,-961){\makebox(0,0)[b]{\smash{{\SetFigFont{12}{14.4}{\rmdefault}{\mddefault}{\updefault}{\color[rgb]{0,0,0}$(0, 0, 0)$}%
}}}}
\put(3556,-961){\makebox(0,0)[b]{\smash{{\SetFigFont{12}{14.4}{\rmdefault}{\mddefault}{\updefault}{\color[rgb]{0,0,0}$(0, 0, 1)$}%
}}}}
\put(3556,-1276){\makebox(0,0)[b]{\smash{{\SetFigFont{12}{14.4}{\rmdefault}{\mddefault}{\updefault}{\color[rgb]{0,0,0}$(2, 0, 0)$}%
}}}}
\put(4951,-961){\makebox(0,0)[b]{\smash{{\SetFigFont{12}{14.4}{\rmdefault}{\mddefault}{\updefault}{\color[rgb]{0,0,0}$(1, 0, 1)$}%
}}}}
\put(4951,-691){\makebox(0,0)[b]{\smash{{\SetFigFont{12}{14.4}{\rmdefault}{\mddefault}{\updefault}{\color[rgb]{0,0,0}$(0, 1, 1)$}%
}}}}
\put(4951,-151){\makebox(0,0)[b]{\smash{{\SetFigFont{12}{14.4}{\rmdefault}{\mddefault}{\updefault}{\color[rgb]{0,0,0}$(0, E, 1)$}%
}}}}
\put(4951,-1276){\makebox(0,0)[b]{\smash{{\SetFigFont{12}{14.4}{\rmdefault}{\mddefault}{\updefault}{\color[rgb]{0,0,0}$(2, 0, 1)$}%
}}}}
\put(4951,-421){\makebox(0,0)[b]{\smash{{\SetFigFont{12}{14.4}{\rmdefault}{\mddefault}{\updefault}{\color[rgb]{0,0,0}$(0, 2, 1)$}%
}}}}
\put(6301,-151){\makebox(0,0)[b]{\smash{{\SetFigFont{12}{14.4}{\rmdefault}{\mddefault}{\updefault}{\color[rgb]{0,0,0}$(2, E, 1)$}%
}}}}
\put(6301,-961){\makebox(0,0)[b]{\smash{{\SetFigFont{12}{14.4}{\rmdefault}{\mddefault}{\updefault}{\color[rgb]{0,0,0}$(1, 2, 1)$}%
}}}}
\put(6301,-556){\makebox(0,0)[b]{\smash{{\SetFigFont{12}{14.4}{\rmdefault}{\mddefault}{\updefault}{\color[rgb]{0,0,0}$(2,1,1)$}%
}}}}
\end{picture}%
}  \hfill  {}
  % \end{center}
  \smallskip
  
  \noindent
  Each of these traces is also a possible trace of the client when it uses
  the abstract object.  \hfill \qed
\end{example}
The next lemma states that when a concurrent object is linearizable
with respect to an abstract object, then it also guarantees contextual
trace refinement of clients that use it.

\begin{lemma}
  Suppose $N$ is a client object, and $OA$ and $OC$ are concurrent objects
  such that $OC$ is linearizable with respect to $OA$. Then $N[OA]
  \sref N[OC]$ holds.
\end{lemma}

\section{Conclusions}

In this paper, we have set up a framework for \emph{studying the links
  between different correctness conditions for concurrent objects and contextual
  trace refinement}, which generalises the results of Filipovi\'{c} et
al. \cite{FORY10}. % The modular nature of our framework can be used to
% study consistency conditions other than linearizability \cite{HeSh08},
% including those for relaxed memory models \cite{DDGS15-ECOOP}. 
We study sequential consistency and linearizability, and
show that sequential consistency does not ensure contextual trace
refinement. We have also shown that linearizability between an abstract
specification and its linearizable implementation implies contextual
trace refinement. % Further work will extend
% \refthm{thm:otr-lin} to arbitrary linearizable implementations, and
% here we expect to build on existing results by Doherty et
% al. \cite{Doherty03,CDG05}. The general nature of our framework will
% also allow us to develop similar results for weak memory
% implementations (e.g., TSO \cite{DDGS15-ECOOP}).

Gotsman and Yang \cite{GY11} also extend Filipovi\'{c} et al.'s work, 
treating non-termination as abort, and considering both safety and progress
properties (lock-freedom). Our trace-based framework can \emph{distinguish}
between non-terminating and aborting programs \cite{HDM10}, and hence is
more general than Gotsman and Yang \cite{GY11}; though liveness properties
are to be considered in future work. 

\bibliographystyle{eptcs}

\bibliography{main,ls}

\end{document}